\documentclass[]{iopart}

\usepackage{graphicx}
\usepackage{mathrsfs} 
\usepackage{color} 

\begin{document}

\title{Determining surface properties with bimodal and multimodal AFM}
\author{Daniel Forchheimer$^1$, Stanislav S. Borysov$^2$, Daniel Platz$^{1,3}$, David B. Haviland$^1$}
\address{$^1$ Nanostructure Physics, KTH Royal Institute of Technology, Roslagstullsbacken 21, SE-106 91, Stockholm, Sweden}
\address{$^2$ Nordita, KTH Royal Institute of Technology and Stockholm University, Roslagstullsbacken 23, SE-106 91, Stockholm, Sweden}
\address{$^3$ Max-Planck-Institute for the Physics of Complex Systems, N\"othnitzer Str. 38, D-01187 Dresden, Germany}

\ead{forchheimer@kth.se}

\begin{abstract}
Conventional dynamic atomic force microscopy (AFM) can be extended to bimodal and multimodal AFM in which the cantilever is simultaneously excited at two ore more resonance frequencies. Such excitation schemes result in one additional amplitude and phase images for each driven resonance, and potentially convey more information about the surface under investigation. Here we present a theoretical basis for using this information to approximate the parameters of a tip-surface interaction model. The theory is verified by simulations with added noise corresponding to room-temperature measurements.
\end{abstract}


\maketitle

\section{Introduction}

The atomic force microscope (AFM) \cite{Binnig1986} is a versatile tool for imaging and characterization of surfaces at the nano and micro meter scale. So-called force spectroscopy is commonly used to obtain the quasi-static force between the tip and the surface as a function of their separation \cite{Cappella1999}, revealing  mechanical and chemical surface properties. Quasi-static measurements are necessarily slow and not conducive to high resolution imaging. Imagining modes such as amplitude modulated AFM (AM-AFM) \cite{Garcia2010} are fast (a few milliseconds per pixel), but the information obtained at each pixel is limited: the amplitude and phase of the response at the drive frequency is not sufficient for quantitative reconstruction of the tip-surface force\cite{Paulo2002}. In this paper we show that it is possible to accurately approximate the tip-surface force while imaging in so-called bimodal and multimodal AFM, without any loss in imaging speed compared to AM-AFM.

Bimodal AFM excites the cantilever at the resonance frequencies of two flexural  eigenmodes of the cantilever \cite{Rodrguez2004,Proksch2006,Martinez2006}. In comparison to AM-AFM,  bimodal AFM provides twice the amount of data (two amplitudes and two phase values) at each image pixel. Bimodal AFM has been demonstrated to increase material contrast \cite{Martinez2008}, and it can be used to quantitatively separate topography from long range force, such as magnetic force \cite{Li2009}. Measurement schemes in AFM have recently been extended to simultaneous excitation and measurement at multiple frequencies \cite{Garcia2012} including: a continuous band \cite{Jesse2007} and discrete comb \cite{Platz2008,Platz2013a,Platz2013} of frequencies around one eigenmode, multiple harmonics of a single drive frequency \cite{Stark2002,Legleiter2006,Sahin2007,Raman2011}, and excitation of more than two eigenmodes \cite{Solares2010} (here denoted multimodal). Despite these advancements, a general framework has been lacking for interpreting the additional signals provided by bimodal and multimodal AFM and relating them quantitatively to the tip-surface force.

A very recent method was presented to analytically calculate parameters of a specific tip-surface force model from the resonant frequency shift of the two eigenmodes, under the conditions of constant response amplitude\cite{Herruzo2014}. Multiple feedback loops are required to keep the response phase and amplitudes constant at each frequency. We take a more general approach and present a method to estimate the tip-surface force directly from the measured amplitudes and phases in open-loop with constant drive conditions.  Our method greatly simplifies the experiment, removes unknown feedback dynamics, and potentially reduces noise. The method is easily extended to arbitrary force models, arbitrarily many excited eigenmodes, and can easily incorporate response at mixing frequencies which occurs off resonance.  The method is an extension of our previous work on the analysis of multi-frequency response in so-called Intermodulation AFM, where a single eigenmode is excited with two closely spaced drive tones and multiple intermodulation products (mixing products) are measured in the response spectrum \cite{Platz2008,Platz2013a,Platz2013}.  Assuming a parametrized model for the tip-surface force, one can fit the spectrum obtained from the model to the measured spectrum and thus obtain a good approximation of the model parameters \cite{Forchheimer2012}. Here we present simulations of a high-Q AFM cantilever and demonstrate accurate reconstruction of tip-surface force when exciting either two or four modes of the cantilever. We investigate the effect of adding noise to the simulation corresponding to a realistic AFM measurement at room temperature in air.  We conclude with a discussion regarding application of the method on experimental data.

\section{Methods}
\subsection{Theory}
We model the multimodal AFM cantilever in the standard way, as a system of coupled harmonic oscillators, each driven with an external drive force \cite{Lozano2008,Kiracofe2013,Borysov2013}. The equation of motion is, 
\begin{equation}
\frac{\ddot q_i}{\omega_i^2} + \frac{\dot q_i}{\omega_i Q_i} + q_i = \frac{1}{k_i} \left( F_{i}(t) + F_{\mathrm{TS}}(d) \right)
\label{eqn:motion}
\end{equation}
where $q_i$, $\omega_i$, $Q_i$, $k_i$ and $F_i$ denote tip-deflection, resonance frequency, quality factor, mode stiffness and the drive force of each mode $i = 1...N$ respectively. The modes are coupled through the tip-surface force $F_{\mathrm{TS}}$ which depends only on the deflection of the tip,
\begin{equation}
d = \sum_{i=1}^{N}q_i.
\label{eqn:deflection}
\end{equation}
Typically the eigen-coordinates $q_i$ can not be independently measured and only the total tip deflection is detected. Indeed, since the tip-surface force depends only on their sum, the system (\ref{eqn:motion}) reduces to a single equation (see Borysov et al. \cite{Borysov2013} for full derivation)
\begin{equation}
d(t) = \chi \left[  F_{drive}(t) + F_{\mathrm{TS}}(d) \right] 
\label{eqn:motion_combined}
\end{equation}
with a linear operator $\chi$ acting on the applied force 
\begin{equation}
\chi =  \sum_{i=1}^{N} \chi_i  = \sum_{i=1}^N \frac{1}{k_i} \left( \frac{1}{\omega_i^2}\frac{d^2}{dt^2} + \frac{1}{\omega_iQ_i}\frac{d}{dt} + 1 \right)^{-1}
\label{eqn:G}
\end{equation}
Furthermore, we assume that the system is weakly nonlinear such that $d(t)$ is periodic with the same period as the drive force $F_{drive}(t)$. Care must be taken to ensure that period doubling or chaotic motion do not occur through proper choice of AFM parameters (set-point, amplitudes etc.) and analysis of multiples of the expected period \cite{Jamitzky2006,Stark2009}. This assumption allows us to measure one period $T$ of the response and express the spectrum of the motion $\hat d(\omega)$ and the spectrum of the tip-surface force $\hat F_{\mathrm{TS}}(\omega)$ as Fourier sums over $\Delta \omega = 2\pi/T$. The discretization of the problem is especially useful because in the frequency domain, the individual linear operators $\hat \chi_i(\omega)$ become arithmetic expressions and $\hat \chi(\omega)$ can be readily expressed as
\begin{equation}
\hat \chi(\omega) =   \sum_{i=1}^{N} \frac{1}{k_i} \frac{1}{-\frac{\omega^2}{\omega_i^2} + i\frac{\omega}{\omega_iQ_i} + 1}.
\label{eqn:Ghat}
\end{equation}

In experiments the drive force can be calibrated by measuring the response far from the surface where the tip-surface force is negligible , $\hat F_{drive}(\omega) = \hat \chi^{-1}(\omega) \hat d_{\mathrm{free}}(\omega)$. Together with (\ref{eqn:motion_combined}) one can now solve for the spectrum of the tip-surface force given the free and engaged motion
\begin{equation}
\hat F_{\mathrm{TS}}(\omega) = \hat \chi^{-1}(\omega) \left( \hat d(\omega) - \hat d_{\mathrm{free}}(\omega) \right).
\label{eqn:Fts}
\end{equation}
If the motion could be accurately measured over a wide band of frequencies, the tip-surface $\hat F_{\mathrm{TS}}(\omega)$ would thus be known. In an AFM experiment detector noise and difficulty in accurately finding the zeros of the transfer function make wide-band measurement impossible.  The force can only be accurately determined at frequencies near a resonance, where the gain is large so that the motion amplitude is significantly larger than the noise floor. Thus, $\hat F_{\mathrm{TS}}$ is known only in a small subset of frequencies, which we call the partial spectrum $\hat F_{\mathrm{TS,partial}}$. The motion is strictly also only known in this partial spectrum, but in contrast to the force,  the partial motion is a good approximation of the true motion. 

In the case of bimodal and multimodal AFM the cantilever is driven with one tone at the resonance frequency of two or more modes. For high Q resonance, the cantilever response to the multifrequency drive dominantly occur at the drive frequencies, as their harmonics and mixing products are not close to a resonance.  Thus the measurable partial spectrum is
\begin{equation}
\hat d \approx \hat d_{\mathrm{partial}}(\omega) = \left\lbrace \begin{array}{ll}
\hat d(\omega) & \omega \in \lbrace \omega_1 ... \omega_M \rbrace \\
0 & \mathrm{else,}
\end{array}
\right.
\label{eqn:dpart}
\end{equation}
where $\lbrace \omega_1 ... \omega_M \rbrace$ is the set of $M$ drive frequencies. Note that there is no requirement that only the drive frequencies be included in the partial spectrum. Any tone which produces a measurable response with a good signal-to-noise ratio can and should be included in $\hat d_{\mathrm{partial}}$ . For the sake of simplicity and to be consistent with most publications on bimodal AFM, we here include response only at driven frequencies in $\hat d_{\mathrm{partial}}$.

\begin{figure}
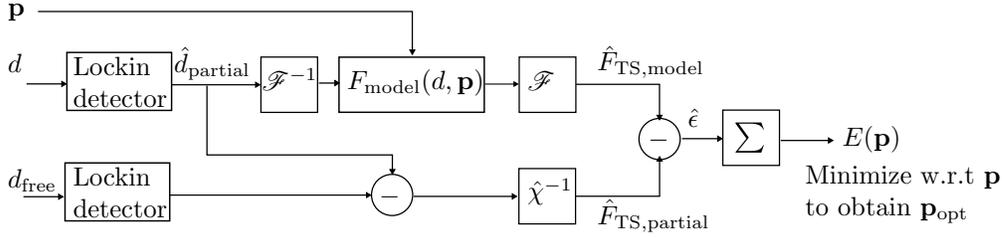
\caption{Flow diagram of the force reconstruction algorithm}
\label{fig:diagram}
\end{figure}

In order to estimate the full tip-surface force from the partial spectrum we follow the approach of Ref~\cite{Forchheimer2012} and introduce a model force $F_{\mathrm{TS}} = F_{\mathrm{model}}(d;\mathbf{p})$, where $\mathbf{p}$ is a vector of model parameters. Evaluating the model with the partial deflection in the time domain as input, we apply the Fourier transform to obtain a parameter-dependent modeled force spectrum $\hat F_{\mathrm{TS,model}}$. Subtracting this modeled force spectrum from the force spectrum calculated directly from the measured partial motion using (\ref{eqn:Fts}), we obtain a frequency and parameter dependent residual $\hat \epsilon$
\begin{equation}
\underbrace{\mathscr{F} \lbrace F_{\mathrm{model}}(d_{\mathrm{partial}}(t);\mathbf{p}) \rbrace}_{\hat F_{\mathrm{TS,model}}(\omega;\mathbf{p})} - \underbrace{ \hat \chi^{-1}(\omega) \left( \hat d_{\mathrm{partial}}(\omega) - \hat d_{\mathrm{free}}(\omega) \right)}_{\hat F_{\mathrm{TS,partial}}(\omega)} = \hat \epsilon(\omega; \mathbf{p}),
\label{eqn:residual}
\end{equation}
where $\mathscr{F}$ denotes the Fourier transform operator. The residual $\epsilon$ is defined only for frequencies included in the partial spectrum so the total error $E(\mathbf{p})$ is the sum of the residual over only these frequencies,
\begin{equation}
E(\mathbf p) = \sum_{i=1}^M \mathrm{Re}[\hat \epsilon(\omega_i; \mathbf{p})]^2 + \mathrm{Im}[\hat \epsilon(\omega_i; \mathbf{p})]^2.
\label{eqn:error}
\end{equation}
If the true motion is used in (\ref{eqn:error}) and $\hat F_{\mathrm{TS,model}}(\omega;\mathbf{p_\mathrm{true}})$ is the true force, this error will be zero for $\mathbf p = \mathbf{p_\mathrm{true}}$ (equation~\ref{eqn:residual} then becomes equation~\ref{eqn:Fts}).  $E$ can only be positive so there will be a minimum at $\mathbf p = \mathbf{p_\mathrm{true}}$ as deviations in $\mathbf p$ means that the model no longer describes the true force. Furthermore we hypothesize that if the partial motion is a good approximation of the true motion, there will be a minimum in $E$ near $\mathbf{p_\mathrm{true}}$
\begin{equation}
\mathbf{p_\mathrm{true}} \approx \mathbf p_{\mathrm{opt}} =  \arg \min_\mathbf{p} E(\mathbf p),
\label{eqn:popt}
\end{equation}
and $\mathbf p_{\mathrm{opt}}$ represents the best fit of the given model to the true tip-surface force.  A flow diagram of the method is presented in Figure~\ref{fig:diagram}.

\subsection{Simulation}
We simulated (\ref{eqn:motion_combined}) using \emph{VODE}, a variable step ODE integrator provided through the Python module \emph{scipy.integrate.ode} version $0.13.3$.  We used realistic AFM cantilever parameters adapted from Ref~\cite{Kiracofe2013} in which the first 4 eigenmodes of an Olympus AC200 cantilever were calibrated (see Table~\ref{tbl:cantilever}). The tip-surface force was modeled  with a repulsive contact force (Hertz model) in one case, and in the other case with an additional attractive van der Waals force (DMT model) as described below.

\begin{table}
\centering
\begin{tabular}{c|c|c|c}
\hline
$i$ & $f_i$ (kHz) & $Q_i$ & $k_i$ (N/m) \\
\hline
1 & 117 & 212 & 4 \\
2 & 674 & 457 & 78.5 \\
3 & 1758 & 507 & 366 \\
4 & 3235 & 600 & 1330 \\
\hline
\end{tabular}
\caption{Cantilever properties for the different bending mode numbers ($i$) used in simulation. The values for resonance frequency ($f_i$), quality factor ($Q_i$) and mode stiffness ($k_i$) have been adapted from Ref \cite{Kiracofe2013}.}
\label{tbl:cantilever}
\end{table}

To avoid Fourier leakage, all drive frequencies and the sampling frequency are chosen to be integer multiples of the measurement bandwidth $\Delta f = 500$~Hz.  The corresponding measurement time $T=1/\Delta f = 2$~ms was longer than the decay-time of the slowest eigenmode, $Q_1 / \pi f_1 = 0.6$~ms.  Transients were avoided by simulating the motion for 20~ms and analyzing only the last 2~ms.  To avoid aliasing in the discrete Fourier analysis, the motion was evaluated with a time step corresponding to 200~MHz, well above the resonance frequency of the highest eigenmode used in the simulation.   

The drive force  $F_{\mathrm{drive}}=\sum_{i=1}^M A_i cos(\omega_i t)$ was chosen with $M=2$ or 4 frequency components, and the amplitude at each frequency was chosen so that the free oscillation amplitude was either equal at all drive frequencies (\emph{equal amplitude scheme}),
\begin{equation}
|\hat d_{\mathrm{free}}(\omega_1)| =|\hat  d_{\mathrm{free}}(\omega_2)| = ...
\label{eqn:equal_amp}
\end{equation}
or such that the stored energy in the oscillation at each eigenmode was roughly equal (\emph{equal energy scheme}), 
\begin{equation}
k_1 |\hat d_{\mathrm{free}}(\omega_1)|^2 = k_2 |\hat d_{\mathrm{free}}(\omega_2)|^2 = ...
\label{eqn:equal_energy}
\end{equation}
In all simulations the sum of the free oscillation amplitudes was adjusted so that the maximum peak-to-peak deflection was 100~nm.

\begin{figure}
\includegraphics[]{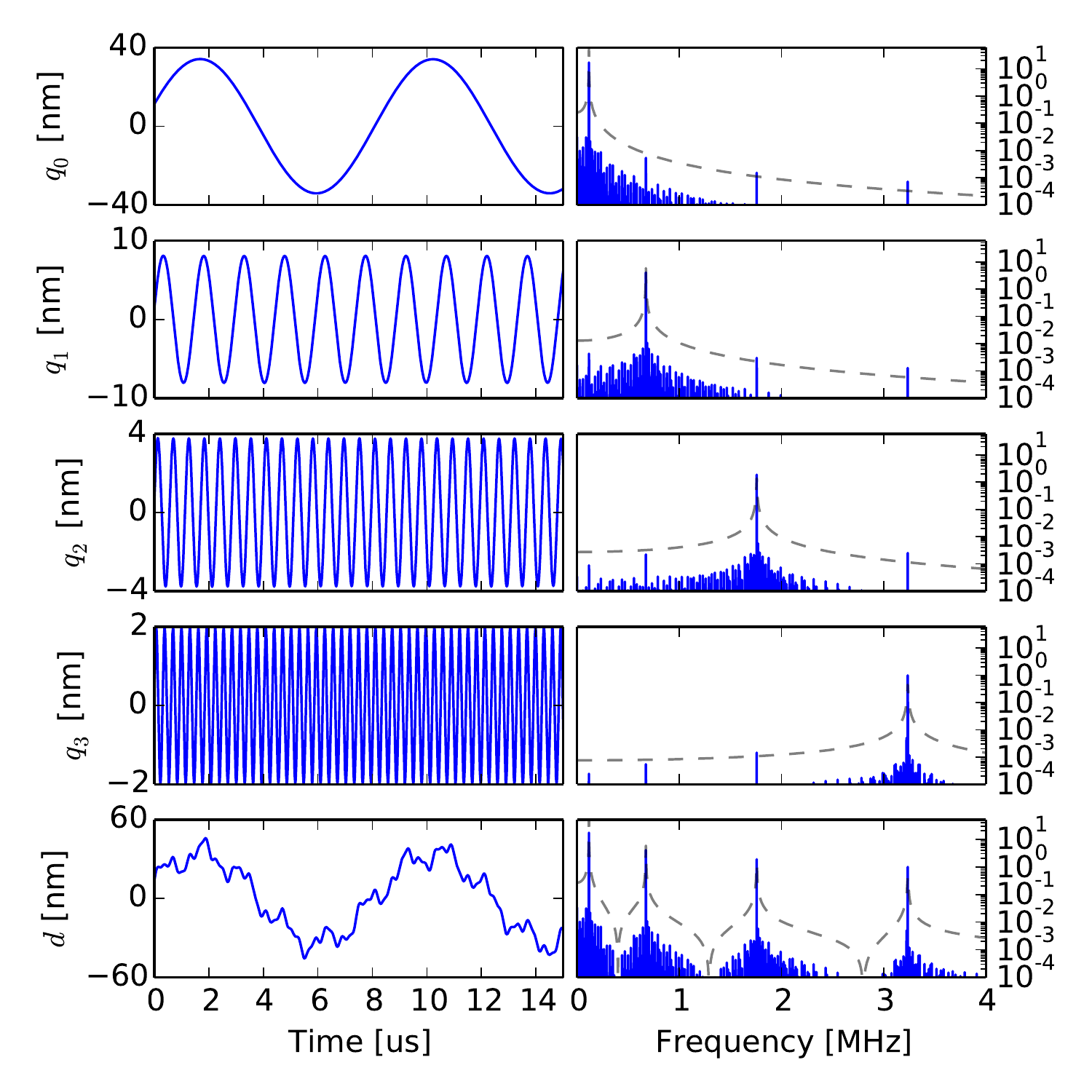}
\caption{Left column: simulated motion of four eigenmodes and the sum tip deflection. The system was driven in the \emph{equal energy scheme} using the DMT force with the surface positioned at $d=-40$~nm. The time domain (left column) shows only 15~$\mu$s of the 2~ms window analyzed for the frequency domain (right column). The dashed gray curve denotes absolute value of the linear transfer function (arb. units) of each individual mode as well as the combined linear transfer function.}
\label{fig:motion}
\end{figure}

Figure~\ref{fig:motion} shows the result of a simulation with four drive tones using the \emph{equal energy scheme} and the DMT model. The motion of each individual eigenmode $q_1 ... q_4$ is plotted in the time and frequency domain. In the spectra one can see that each mode has strongest response at the four drive frequencies. We also see the presence of many additional tones which are mixing tones or intermodulation products of the four drive tones. This dense comb of response at many frequencies stems from the fact that the system is nonlinear, and that the drive frequencies are not harmonics (integer multiples) of the lowest drive frequency.

\begin{figure}
\includegraphics[]{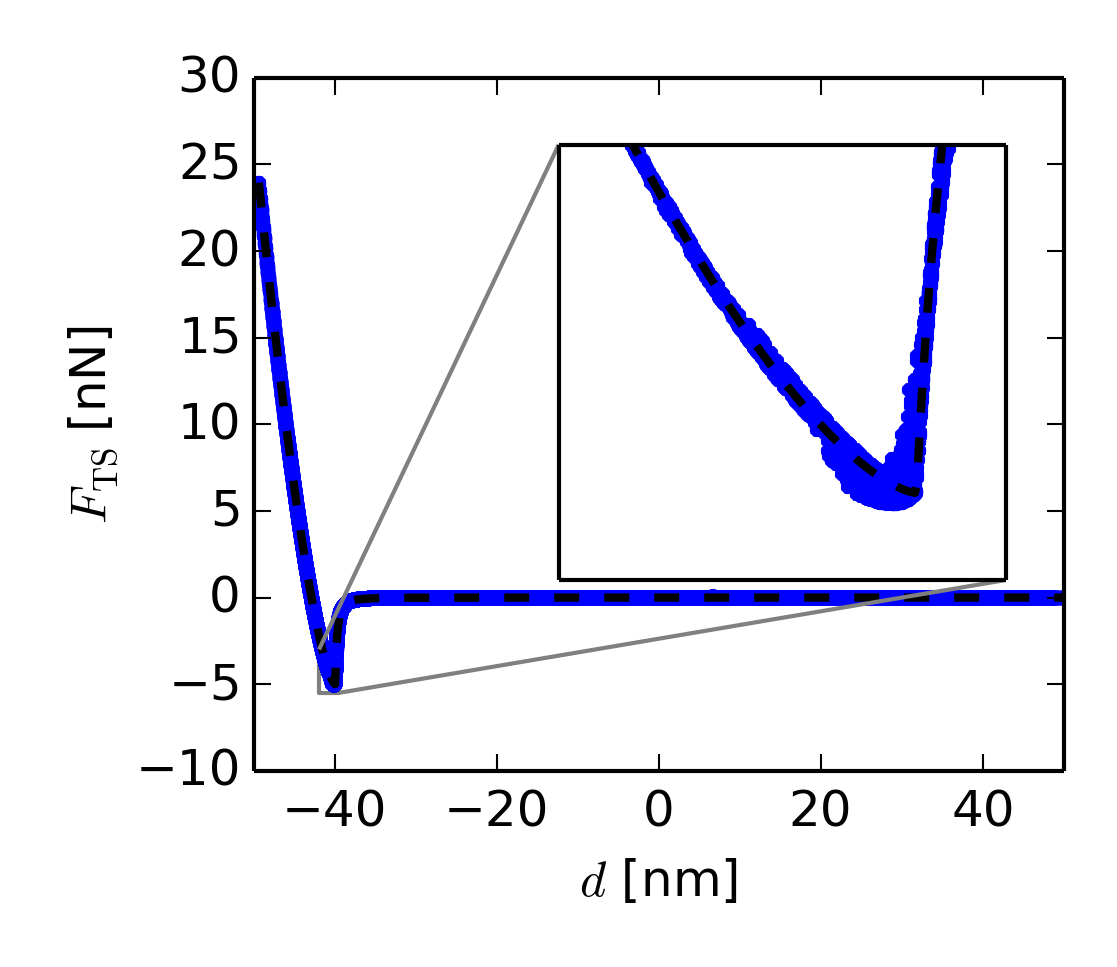}
\caption{Test of the accuracy of the simulator.  Excellent agreement between the tip-surface force used in the simulation (dashed black line) and the tip-surface force calculated from the inverse Fourier transform of (\ref{eqn:Fts}) with the full broad-band response spectrum of $\hat{d}$. }
\label{fig:sim_accuracy}
\end{figure}

To test the accuracy of the simulation we calculated $\hat F_{\mathrm{TS}}(\omega)$ from (\ref{eqn:Fts}) using the entire response spectrum $\hat d(\omega)$. We then used the inverse Fourier transform to obtain $d(t)$ and $F_{\mathrm{TS}}(t)$, and plot them against each other to obtain the tip-surface force curve $F_{\mathrm{TS}}(d)$. This force curve was in excellent agreement with the actual force used in the simulation (see Figure~\ref{fig:sim_accuracy}), with a maximum deviation of 800~pN and standard deviation of 10~pN.  We conclude that the numerical error in the simulator is small.

\subsection{Adding noise to simulation}
There are typically two major noise contributions in AFM: the detector noise which gives an equivalent deflection noise that is frequency independent (white noise), and the thermal noise which is a white force noise driving the system, coloured in the deflection signal \cite{Hutter1993}. Close to a resonance, as for frequencies analysed in this paper, force noise dominate over detector noise and in this case the measurement is at a fundamental thermal limit of sensitivity. The magnitude of the thermal force noise is found from the fluctuation-dissipation theorem, which gives the single-sided power spectral density of motion fluctuation at each eigenmode \cite{Saulson1990}
\begin{equation}
S_{q_iq_i} = -\frac{2k_BT}{\pi f_i} \mathrm{Im}[ \hat \chi_i ] = 2k_BT\frac{k_i}{\pi f_i Q_i }|\hat \chi_i |^2,
\end{equation}
where $k_B$ is the Boltzmann constant and T the temperature. Thus, the system is excited by a frequency independent noise force of magnitude
\begin{equation}
F_{\mathrm{noise},i} = \sqrt{\frac{2 k_B Tk_i}{ \pi f_i Q_i}}.
\end{equation}
For each mode of the simulated cantilever, the room temperature power density of the noise force was: 21, 26, 33 and 43 fN/Hz$^{1/2}$ respectively. For the simulated measurement bandwidth of $\Delta f = 500$~Hz this gave a standard deviation of the  motion fluctuation, $\Delta \hat q_i = \frac{Q_i}{k_i} F_{\mathrm{noise},i} \sqrt{\Delta f}$, of 24.4, 3.4, 1.0 and 0.4~pm respectively.

To properly account for the thermal noise force, one should add a random force in each time step of the numerical integration of (\ref{eqn:motion}). However, as the noise force was small compared to the drive force, and for simplicity, we simulated the noise-free nonlinear response, and subsequently added the noise prior to analysis. Gaussian noise with a standard deviation of $\Delta \hat q_i/\sqrt{2}$ was added to each of the complex quadratures in the frequency domain. We assume that this equilibrium, linear response approach will slightly under-estimate the magnitude of the noise and its negative effects on reconstruction of the force curves. However, this simplistic approach did allow for fast computation.

\section{Results and Discussion}

\subsection{Two parameters: Hertzian tip-surface force}
With bimodal AFM (driving at two eigenmodes) it is possible to reconstruct a force described by a model with only two free parameters.  The model we have choosen is the Hertz model of contact mechanics from the late 1800's \cite{Hertz1896}.  This model accounts for the repulsive forces due to the mutual deformation of two elastic bodies in contact.  The Hertz model neglects adhesion. Given the geometry of a spherical tip indenting a flat surface the Hertz tip-surface force can be written as \cite{Kiracofe2012}
\begin{equation}
F_{\mathrm{TS}}(d) = \left\lbrace
\begin{array}{ll}
0 & d>p_0 \\
p_1 (p_0-d)^{3/2} & d \leq p_0 \\
\end{array}
\right. 
\label{eqn:hertz}
\end{equation}
where $p_0$ is the position of the surface in the coordinate of the tip deflection and $p_1$ is the "strength" of the interaction. The later parameter is more commonly expressed as $p_1 = \frac{4}{3}E^\star \sqrt{R}$ where $R$ is the tip radius and $E^\star$ the effective elastic modulus $E^\star = \left( (1-\nu_{tip}^2)/E_{tip} + (1-\nu_{surface}^2)/E_{surface} \right)^{-1}$ with the $E$'s and $\nu$'s being the elastic moduli and Poisson ratios of the tip and surface materials respectively.  It is apparent from the equation that one can not obtain the elastic modulus of the surface from a measurement of $F_{\mathrm{TS}}(d)$ alone. Further assumptions or a calibration of the tip radius and the Poisson ratios are required. 

We are interested in investigating the possibility to quantitatively obtain the force-distance dependence from an AFM measurement, which is why we describe the force model with parameters $p_0$ and $p_1$ rather than material properties. The validity of the force model for specific materials are outside the scope of this paper. In the simulation we chose $p_0=-40$~nm and $p_1=1.0$~GPa~nm$^{-1/2}$, which we denote as 'true' values. These parameter values roughly correspond to an AFM experiment with 80\% amplitude set-point on a polymer material.

\begin{figure}
\includegraphics[]{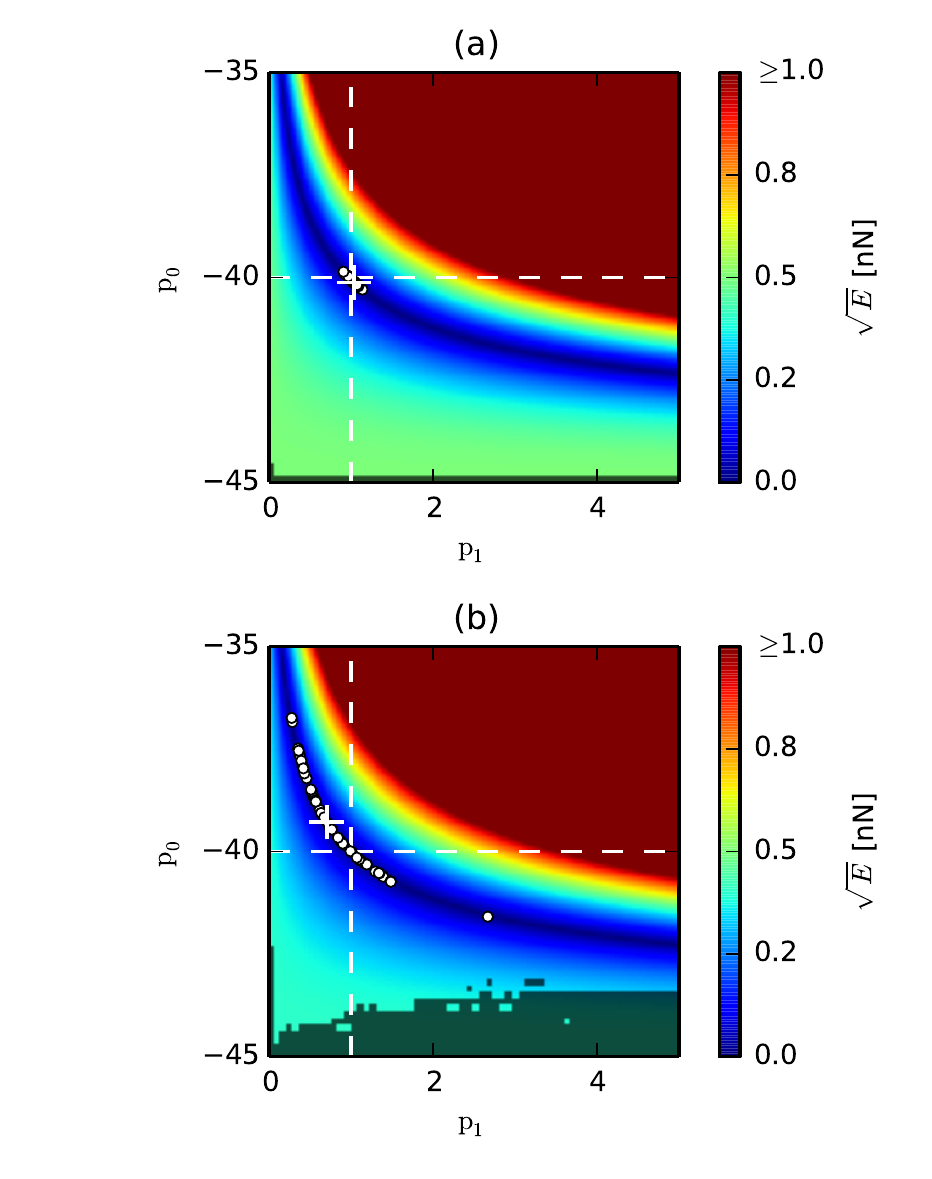}
\caption{Error in the parameter space of the Hertz model. (a) \emph{equal energy scheme} and (b) \emph{equal amplitude scheme}. White dashed lines show the true parameters values used in simulation and the white cross shows the minimum found in the absence of noise. White circles are at minima found with thermal noise. The grey region denotes values of initial parameters for which the solver did not converge.}
\label{fig:error_space}
\end{figure}

From the simulated cantilever motion we filtered out the partial response spectrum $\hat d_{\mathrm{partial}}$ defined as the response only at the two drive frequencies and calculate the error defined by (\ref{eqn:error}), $E(p_0,p_1)$. We hypothesized that there should be a minimum if $p_0$ and $p_1$ were the values used in the simulation, which we tested by calculating $E$ for a range of $p_0$ and $p_1$ around the true value. Figure~\ref{fig:error_space} shows a low value for the error around the expected parameter values. However, the minimum is not well localized in the parameter space, and it forms an extended, curved trench with steep side-walls. Thus it is very hard to judge graphically if the true parameter values represent a minimum in this parameter space.

To further investigate the presence of a minimum in $E$ we used a numerical solver provided in \emph{Scipy.optimize} implementing the Levenberg-Marquardt algorithm, which is well suited for finding a local minimum in nonlinear least-square problems \cite{More1978}. For the \emph{equal energy scheme}, equation (\ref{eqn:equal_energy}), the solver found a minimum very close to the true parameter values, while for the \emph{equal amplitude scheme}, equation  (\ref{eqn:equal_amp}), the minimum found by the solver was slightly off (see Table~\ref{tbl:hertz}). We attribute this systematic error between the optimal parameters found by the solver and the true parameter values to come from the use of only a partial response spectrum $\hat d_{\mathrm{partial}}$ when evaluating the model force.

The numerical solver finds the minimum by iteration from an initial parameter value. For a wide range of initial parameter values we found that the solver converged to the same minimum, with no dependence on the initial values. The initial values for which the solver failed to converge are shaded gray in Figure~\ref{fig:error_space}. Only when $p_0$ was several nm away from the true value did the solver fail, which is in agreement with our previous findings using Intermodulation AFM\cite{Forchheimer2012}. As the maximum oscillation amplitude is directly measured and the surface can be expected to  be close to the turning point, it is possible to provide a good initial estimate of $p_0$ for which the solver will converge to a solution when analyzing experimental data.

\begin{figure}
\includegraphics[]{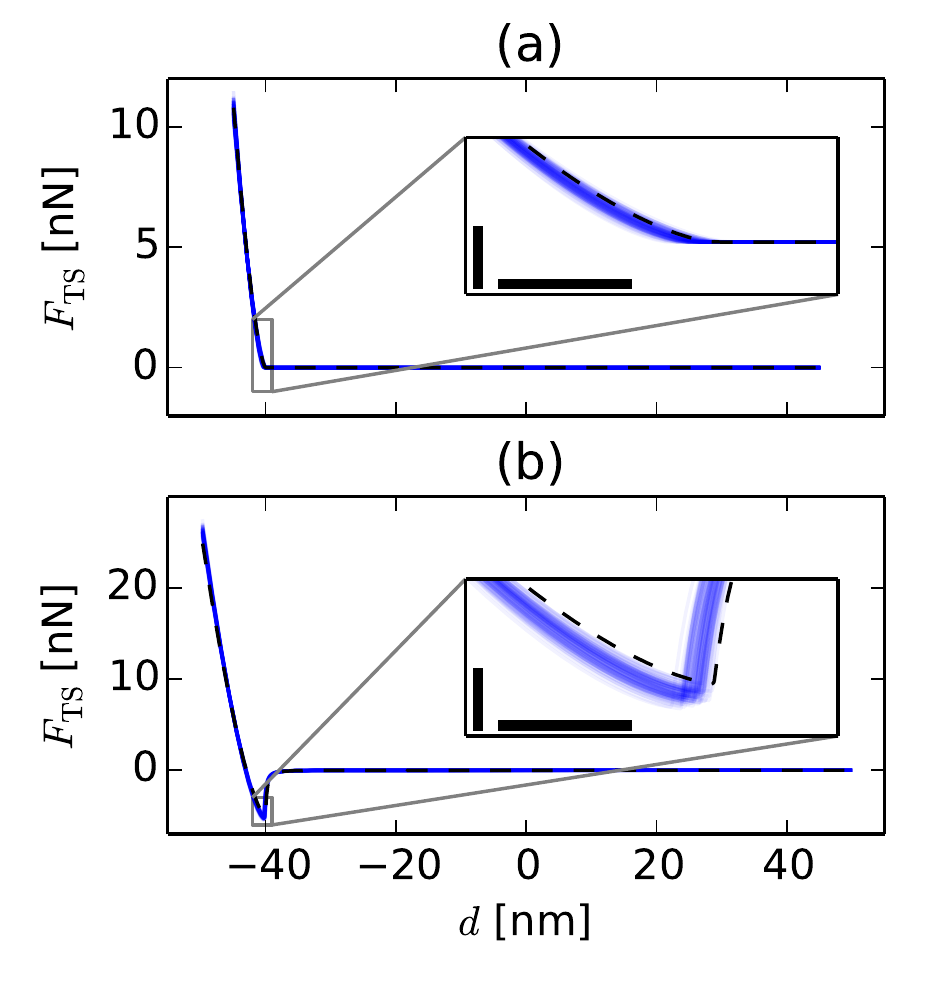}
\caption{Reconstructed force curves. (a) Hertz model reconstructed from bimodal AFM. Simulated force curve (dashed black) and reconstructions with noise (light blue). (b) DMT model reconstructed from four-mode AFM, simulated (dashed black) and reconstructed with noise (light blue). Scale bar in the zoom-insets are 1~nm and 1~nN for the x and y-axis respectively.}
\label{fig:forcecurve}
\end{figure}

To evaluate the effect of noise on the numerical optimization, we added noise to each $\hat q_i$ as described above. We estimated the mean and standard deviation of the fitted parameters from an ensamble of 100 such calculations (Table~\ref{tbl:hertz}). Comparing the two drive schemes we found that the simulation using the \emph{equal amplitude scheme} gave a larger spread in the parameter values than the simulation using the \emph{equal energy scheme}. In both cases the fitted parameters all fall within the elongated minimum found in the parameter space. The shape of elongated, curved minimum tells us that the two parameters are not independent of each other, and therefore a statistically independent mean and standard deviation for $p_0$ and $p_1$ are not good quantities for testing of the accuracy of the method. Rather than standard deviation of each parameter, one should consider the overall accuracy of the tip-surface force curve. Figure~\ref{fig:forcecurve}a shows each of the reconstructed force curves in the presence of noise for the equal energy scheme, where one can see that the deviation from the true force curve is typically less than 1~nN.

We also performed a simulation with the same Hertz model in which all four modes were driven using the \emph{equal energy scheme}. Analyzing response at four frequencies for a two-parameter model means that the reconstruction problem is over-determined. In this case we found that the systematic error was similar to the case of two drive tones, but the sensitivity to noise was reduced.

\begin{table}
\begin{tabular}{r|c|c}
\hline
 & $p_0$ (nm) & $p_1$ (GPa nm$^{-1/2}$) \\
\hline 
True & $-40.00$ & $1.00$\\
&&\\
2 tones, equal energy & $-40.12$ & $1.04$\\
\;with noise & $-40.11 \pm 0.10$ & $1.03 \pm 0.05$\\
&&\\
2 tones, equal amplitude & $-39.28$ & $0.70$\\
\;with noise & $-39.18 \pm 1.07$ & $0.80 \pm 0.45$\\
&&\\
4 tones, equal energy & $-40.12$ & $1.05$\\
\;with noise & $-40.12 \pm 0.05$ & $1.05 \pm 0.02$\\
\hline
\end{tabular} 
\caption{Comparison of the true parameters and extracted parameters for a Hertz contact model. Extracted parameters are presented in the absence of noise and with a thermal noise corresponding to $300$~K. For the parameter values with noise 100 independent fits were performed the mean value is reported with one standard deviation as error margins. }
\label{tbl:hertz}
\end{table}

\subsection{Four parameters: DMT tip-surface force}
Often adhesion between the tip and surface can not be neglected. A model which attempts to account for adhesion and is often used in the AFM literature, is the Derjaguin-Muller-Toporov (DMT) model, where a van-der-Waals attractive force is piece-wise connected to the Hertz model.  For a flat surface and spherical tip this model is\cite{Kiracofe2012}
\begin{equation}
F_{\mathrm{TS}}(d) = \left\lbrace
\begin{array}{ll}
-p_2 \left( \frac{p_3}{d-p_0+p_3} \right)^2 & d>p_0 \\
-p_2 + p_1 (p_0-d)^{3/2} & d \leq p_0 \\
\end{array}
\right. 
\label{eqn:DMT}
\end{equation}
The two additional parameters are the force of adhesion $p_2$ and the finite distance between the surfaces "in contact" $p_3$, typically assumed to be atomic separation. The adhesion force $p_2$ is more commonly expressed in terms of the tip radius $R$ and the Hamaker constant $H$ between the tip and the surface, $p_2= HR/p_3^2$.

This model has four free parameters and response must therefore be measured for at least four frequencies. We excited the cantilever with one drive tone close to each of the four eigenmodes and with the numerical solver we were able to successfully fit the parameter values within 5\% of the true values in the \emph{equal energy scheme}, while the fit error using \emph{equal amplitude scheme} was slightly larger (see Table~\ref{tbl:dmt}). Again we attribute this systematic error with no added noise to result from the approximation that $\hat d_{\mathrm{partial}} \approx \hat d$. Similarly to the Hertz model, we found that the \emph{equal energy scheme} was rather insensitive to noise with deviation of only a few percent, while the \emph{equal amplitude scheme} showed a larger spread in the fitted parameters. Figure~\ref{fig:forcecurve}b shows the reconstructed force distance curves for the \emph{equal energy scheme}, where one can see that the force curve was correctly reconstructed within a few nN.

\begin{table}
\begin{tabular}{r|c|c|c|c}
\hline
 & $p_0$ [nm] & $p_1$ [GPa nm$^{-1/2}$] & $p_2$ [nN] & $p_3$ [nm]\\
\hline
True & $-40.00$ & $1.00$& $5.00$& $0.50$\\
&&&&\\
4 tones, equal energy & $-40.14$ & $1.07 $& $5.23$ & $0.49$\\
\;with noise & $-40.16 \pm 0.07$ & $1.08 \pm 0.03$ & $5.24 \pm 0.11$ & $0.49 \pm 0.04$\\
&&&&\\
4 tones, equal amp & $-40.02$ & $1.07 $& $5.23$ & $0.65$\\
\;with noise & $-39.92 \pm 0.25$ & $1.09 \pm 0.33$ & $5.60 \pm 1.25$ & $0.46 \pm 0.67$\\
\hline
\end{tabular}
\caption{Comparison of the true parameters and extracted parameters for a DMT contact model.}
\label{tbl:dmt}
\end{table}

\subsection{Toward experiments}
Applying the presented method to actual AFM experiments requires accurate calibration of the cantilever mode stiffness; an issue under active investigation \cite{Lozano2010,Borysov2014}. Measurement is further complicated by the fact that the AFM typically does not measure the tip deflection $d$ directly. The most common optical lever technique \cite{Meyer1988} measures a voltage $V$ on the photo detector which is proportional to the angle at the end of the cantilever. For small angles and single eigenmode motion, this angle is linearly proportional to the cantilever deflection and thus the deflection can directly be obtained from the detector voltage $d= \alpha V$ with one linear calibration constant $\alpha$.

However, the different eigenmode-shapes of the cantilever give a different relationship between angle and deflection.  Thus an individual value $\alpha_i$ is required for each mode\cite{Kiracofe2010,Schaffer2005}, further complicating the calibration and determination of the tip motion. Large spot sizes can also be problematic in defining $\alpha_i$ and measuring response from higher eigenmodes \cite{Stark2004a}.

Although the method is demonstrated for two specific force models, we believe the same general method could be applicable to many different models of the tip-surface force. For each material a suitable model would have to be chosen and its validity confirmed. To allow for models with increased number of free parameters, while avoiding the use of higher eigenmodes, one can also increase the number of observables by inclusion of nonlinear mixing products in the partial motion and in  (\ref{eqn:error}).

\section{Conclusions}
Dynamic AFM with multiple flexural eigenmodes gives access to new information channels not accessible with single frequency AM-AFM. Despite this improvement the added information is limited in scope and can not be used to blindly reconstruct a complex tip-surface force. If the tip-surface force is approximated with a model containing only a few free parameters, one can fit these parameters to the measured data and thus obtain an approximation of the tip-surface force. We have shown with simulations that the expected parameter values can be obtained from measurements in realistic conditions . In our simulations we observed that it was advantageous to excite the cantilever so that the energy stored in each mode of the freely oscillating cantilever was equal, as opposed to excitation where the response amplitude for each mode was constant. This finding could be related to previous research where it was found that second mode response was more strongly dependent on the distance to the surface for larger amplitude ratio \cite{Stark2009}. We further found that in order to reduce the variation in tip-surface force parameters resulting from noise, more eigenmodes should be excited and measured than the number of free parameters in the tip-surface force model. This finding motivates an effort to increase the number of eigenmodes used in multimodal AFM. However, there should be an equally strong effort expanded in finding ways to accurately calibrate these additional eigenmodes. 

Provided that accurate calibration of each of the higher eigenmodes can be performed the force reconstruction method presented here should be applicable to experimental data. The method makes use of all the information in the deflection spectrum which can be measured above the noise floor and therefore we consider it to be an optimal method for approximating the tip-surface force in bimodal and multimodal AFM measurements.

\ack
We acknowledge Olle Engkvist Foundation, Knut and Alice Wallenberg Foundation and Vetenskapsr\aa det for financial support.

\bibliographystyle{unsrt}
\bibliography{references}
\end{document}